\documentclass[twocolumn,showpacs,preprintnumbers,amsmath,amssymb]{revtex4}
\usepackage{graphicx}
\usepackage{dcolumn}
\usepackage{bm}
\begin{document}
\title{
Crossover behaviors in liquid region of vortex states\\
above a critical point caused by point defects
}
\author{Yoshihiko Nonomura}
\altaffiliation[Also at ]{Lyman Lab.\ of Physics, 
Harvard University, Cambridge, MA 02138}
\email{nonomura.yoshihiko@nims.go.jp}
\author{Xiao Hu}
\affiliation{
Computational Materials Science Center, National Institute 
for Materials Science, Tsukuba, Ibaraki 305-0047, Japan}
\date{\today}
\begin{abstract}
Vortex states in high-$T_{\rm c}$ superconductors with point 
defects are studied by large-scale Monte Carlo simulations of 
the three-dimensional frustrated XY model.  A critical point 
is observed on the first-order phase boundary between the 
vortex slush and vortex liquid phases. A step-like anomaly of 
the specific heat is detected in simulations of finite systems, 
similar to an experimental observation [F.~Bouquet {\it et al.}, 
Nature (London) {\bf 411}, 448 (2001)]. However, it diminishes 
with increasing system size, and the number and size distribution of 
thermally-excited vortex loops show continuous behaviors around this 
anomaly. Therefore, the present study suggests a crossover rather 
than a thermodynamic phase transition above the critical point.
\end{abstract}
\maketitle
Intensive studies on vortex states of high-$T_{\rm c}$ 
superconductors have revealed complex structures of the 
phase diagram in the magnetic field along the $c$ axis. 
Point defects such as oxygen deficiencies are essential for 
such complexity. The Bragg glass (BrG) \cite{Nattermann,Giamarchi} 
phase is stable at low temperatures and low magnetic 
fields. This phase melts into the vortex liquid (VL) or vortex 
glass (VG) \cite{Fisher} phases as temperature $T$ or 
magnetic field $B$ increases, respectively. The BrG 
phase has a closed first-order boundary in the $B$-$T$ 
phase diagram. \cite{Ertas,Giamarchi2} The vortex 
slush (VS) \cite{Worthington} phase locates between 
the VG and VL phases, and another first-order boundary 
between the VS and VL phases terminates at a critical 
point. \cite{Worthington,Kierfeld} 

The BrG-VL, \cite{Safar92,Kwok94,Zeldov95,Welp96,Schilling96} 
BrG-VG \cite{Cubitt93,Safar95,Khaykovich97,Nishizaki98} 
and VS-VL \cite{Nishizaki00,Wen01} phase boundaries have 
been observed experimentally. Similar phase diagrams have 
been obtained by numerical simulations including point 
defects, \cite{Gingras,Ryu,Otterlo,Sugano,Nonomura} 
though the stability of the VG phase is still 
controversial. \cite{Olsson,Vestergren} 

Recently Bouquet {\it et al.} \cite{Bouquet} observed a 
$\delta$-function peak and a sharp step of the specific 
heat below and above a critical point, respectively. 
They argued that this step-like anomaly corresponds 
to a second-order phase transition related to a 
vortex-loop blowout. \cite{Tesanovic,Nguyen}

\begin{figure}
\includegraphics[width=80mm]{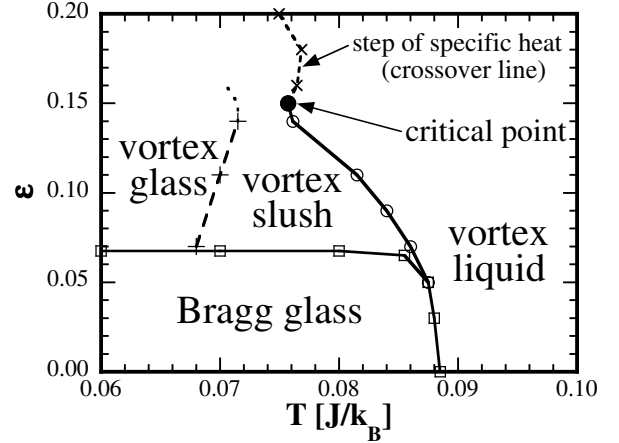}
\caption{\label{fig1}Phase diagram of the model (\ref{XYham}) 
in the plane of the strength of point defects ($\epsilon$) and 
temperature based on a finite system with $L_{a}=L_{b}=50$ and 
$L_{c}=40$. Part of this phase diagram (for $\epsilon \leq 0.14$) 
has already been published. \cite{Nonomura} The first-order 
VS-VL phase boundary terminates at the critical point (full 
circle), and the step of the specific heat (dotted line) is 
observed above the critical point. The VG-VS phase boundary 
for $\epsilon>0.14$ is out of the scope in the present study.
}
\end{figure}

In the present article, we investigate the phase diagram in the 
vicinity of the critical point of the VS-VL phase boundary. 
As displayed in Fig.\ \ref{fig1}, we find that the critical 
point is located at $\epsilon\approx 0.15$, which is just 
above the maximum value of $\epsilon$ in our previous 
work. \cite{Nonomura} A step-like anomaly of the 
specific heat is observed above the critical point in 
finite systems. However, the size dependence of this 
anomaly reveals that this step does not indicate a 
thermodynamic phase transition but a crossover. 

In the present study we use the same model as investigated in our 
previous article on vortex states with point defects, \cite{Nonomura} 
namely the  three-dimensional anisotropic, frustrated XY model 
on a simple cubic lattice, \cite{Li,Hu}
\begin{eqnarray}
  \label{XYham}
  {\cal H}&=&-\hspace{-0.2cm}
              \sum_{i,j \in ab\ {\rm plane}} \hspace{-0.3cm}
                   J_{ij}\cos \left(\phi_{i}-\phi_{j}-A_{ij}
                                    \right)\nonumber\\
          & &-\frac{J}{\Gamma^{2}} \hspace{-0.1cm}
               \sum_{m,n\parallel c\ {\rm axis}}\hspace{-0.2cm}
               \cos \left(\phi_{m}-\phi_{n}\right)\ ,\\
   A_{ij} &=&\frac{2\pi}{\Phi_{0}}\int^{j}_{i}{\bm A}^{(2)}
                                       \cdot {\rm d}{\bm r}^{(2)},\ 
             {\bm B}={\bm \nabla}\times {\bm A}\ ,
\end{eqnarray}
with the periodic boundary condition in all directions. 
A uniform magnetic field ${\bm B}$ is applied along the $c$ 
axis, and the averaged number of flux lines per plaquette is 
$f$. A point defect is introduced as the plaquette surrounded 
by four weaker couplings and distributed randomly in the $ab$ 
plane with probability $p$. Couplings are $J_{ij}=(1-\epsilon)J$ 
($0<\epsilon<1$) on the point defects, and $J_{ij}=J$ elsewhere. 
As in our previous article, we fix the parameters $\Gamma=20$, 
$f=1/25$ and $p=0.003$, and vary the strength of point defects 
$\epsilon$ and temperature $T$. Configurations of defects are 
also the same as in our previous work. Monte Carlo simulations 
are started from a very high temperature, and systems are gradually 
cooled down. Typical Monte Carlo steps are $4\sim 14\times 10^{7}$ 
for equilibration, and $2\sim 16\times 10^{7}$ for measurement. 

\begin{figure}
\includegraphics[width=80mm]{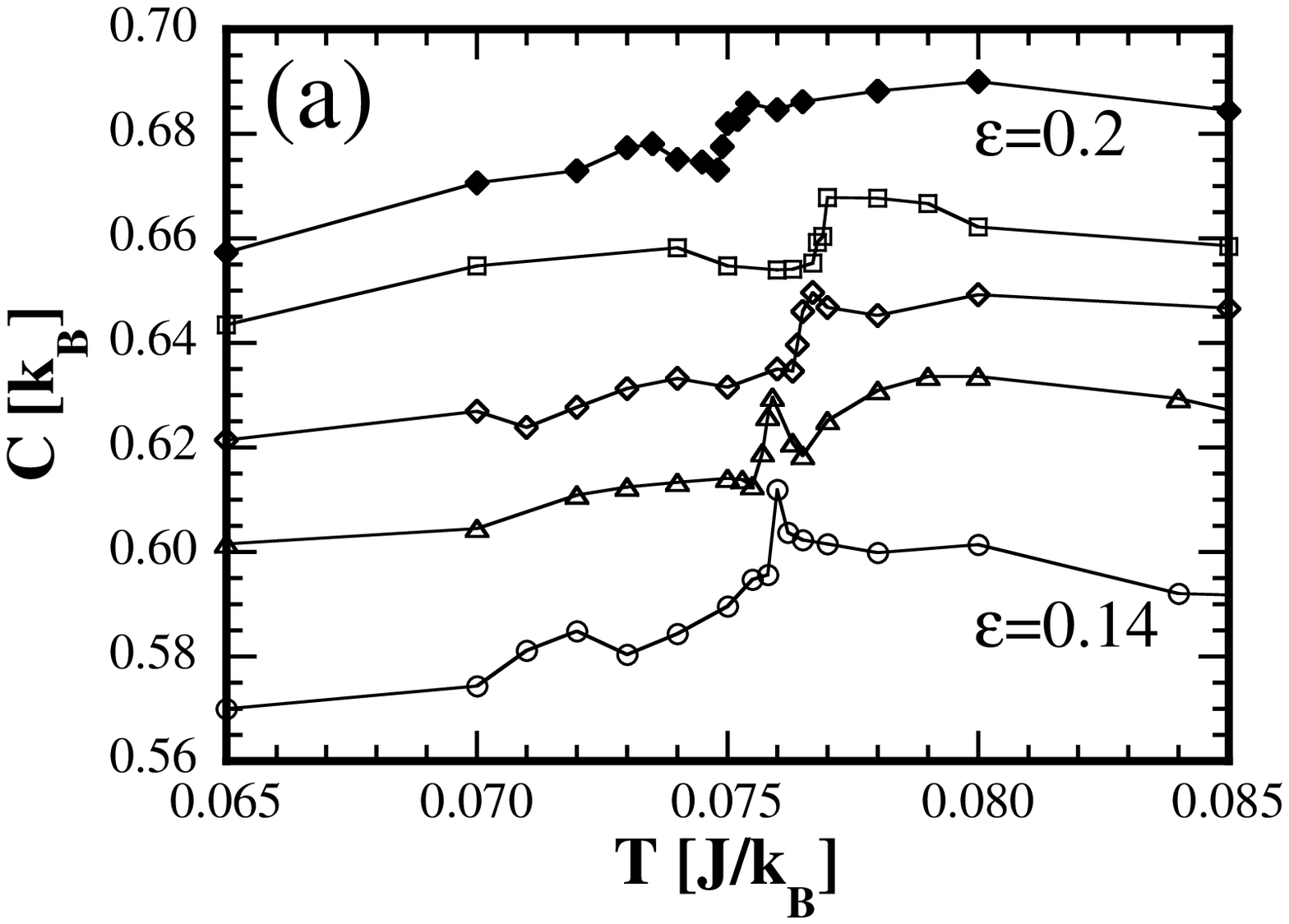}
\includegraphics[width=80mm]{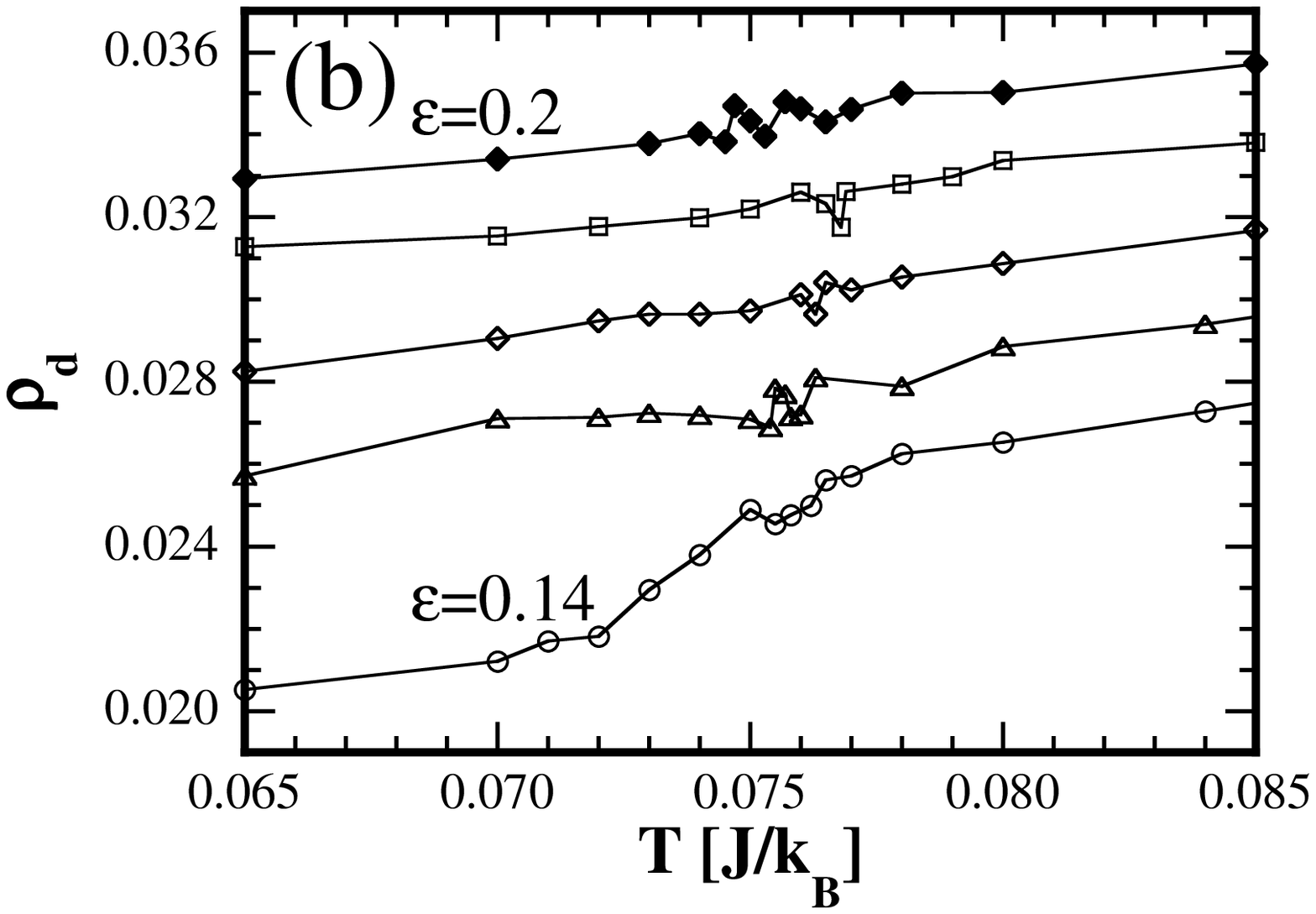}
\includegraphics[width=80mm]{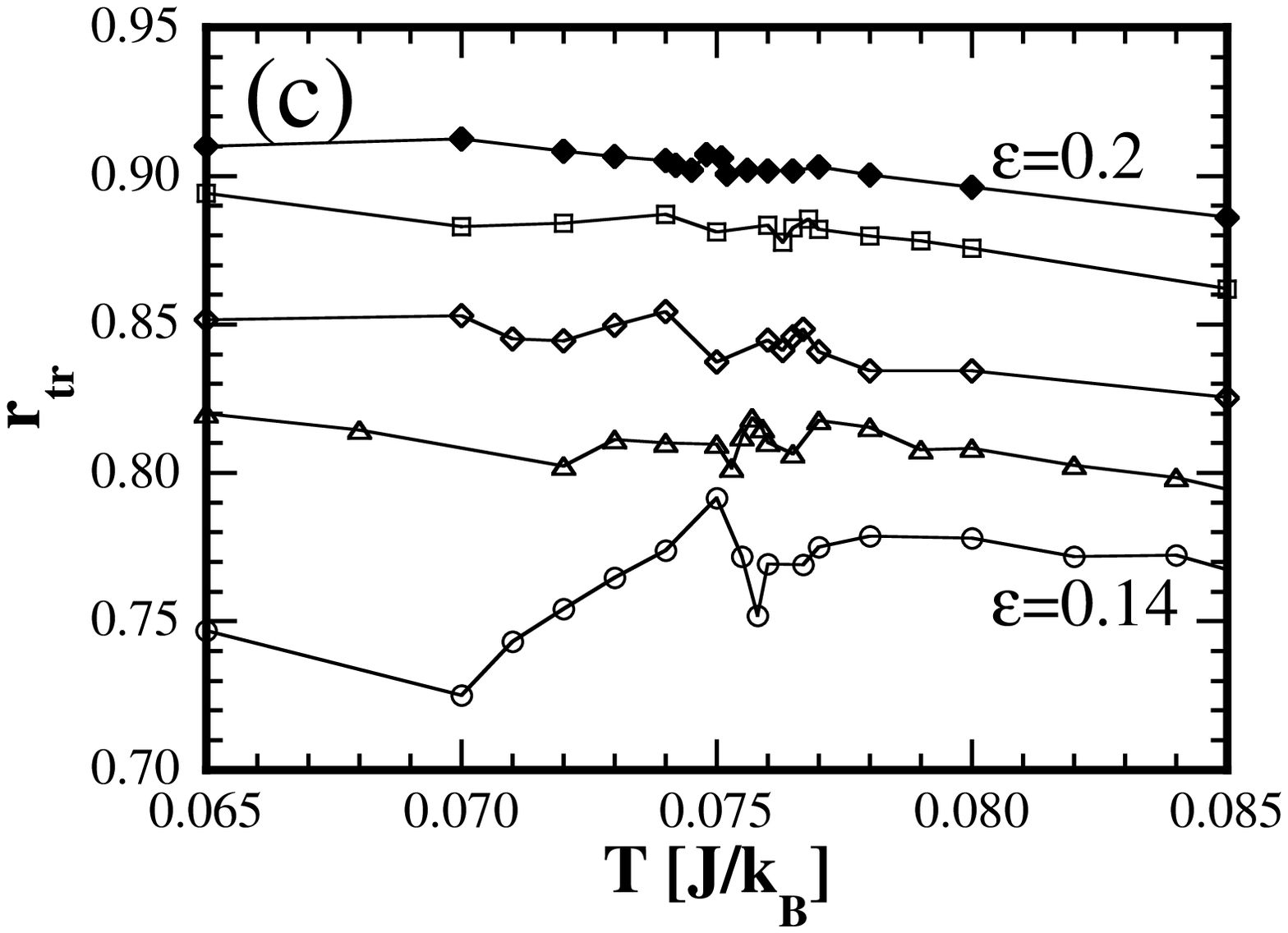}
\caption{\label{fig2}Temperature dependence of (a) the specific 
heat, (b) the density of dislocations in the $ab$ plane and (c) the 
ratio of point defects which trap flux lines for $\epsilon=0.14$, 
$0.15$, $0.16$, $0.18$ and $0.2$ (from bottom to top). The origin 
of the quantities are  varied for each $\epsilon$ in (a) and (b).
}
\end{figure}
First, we concentrate on the system with $L_{a}=L_{b}=50$ 
and $L_{c}=40$. Temperature dependence of the specific 
heat $C$ for $0.14\leq\epsilon\leq 0.2$ is displayed 
in Fig.\ \ref{fig2}(a). At $\epsilon=0.14$, $C$ has a 
$\delta$-function peak at $T_{\rm sl}=0.0760 J/k_{\rm B}$. 
This peak becomes very small at $\epsilon=0.15$, 
and is replaced by a step for $\epsilon\geq 0.16$. 
This result suggests that the critical point locates 
at $\epsilon\approx 0.15$. The phase diagram obtained 
from this finite system is summarized in Fig. \ref{fig1}. 

Consistent behaviors are seen in other quantities. Temperature 
dependence of the density of dislocations in the $ab$ plane 
$\rho_{\rm d}$ and the ratio of point defects which trap 
flux lines $r_{\rm tr}$ are given in Figs.\ \ref{fig2}(b) and 
\ref{fig2}(c), respectively. The quantity $\rho_{\rm d}$ 
characterizes \cite{Kierfeld,Nonomura} the VS-VL phase 
transition. At $\epsilon=0.14$, $\rho_{\rm d}$ jumps 
between $T=0.072 J/k_{\rm B}$ and $0.076 J/k_{\rm B}$, 
where $C$ sharply increases. This jump becomes much smaller 
at $\epsilon=0.15$ and invisible for $\epsilon \geq 0.16$. 
Keeping possible statistical errors in mind, a tiny hump and  dip 
seem to exist at the temperature region $T^{\ast}$ where the 
specific heat has a step-like anomaly for $\epsilon \geq 0.15$, 
but the hump and dip are one order smaller than the jump of 
$\rho_{\rm d}$ related to the VS-VL phase transition, 
and therefore considered to be finite-size effects. 

At $\epsilon=0.14$, the quantity $r_{\rm tr}$ gradually 
increases from $T=0.070 J/k_{\rm B}$ as approaching 
the transition temperature $T_{\rm sl}$, because the 
short-range order of flux lines in the VS phase becomes 
fragile as temperature increases. Then, it rapidly decreases 
just below $T_{\rm sl}$, because the mobility of flux lines 
becomes larger in the coexisting region of the first-order 
phase transition. This anomaly becomes obscure at 
$\epsilon=0.15$ and disappears for $\epsilon\geq 0.16$. 
Although a tiny hump and dip also seem to exist at $T^{\ast}$ 
for $\epsilon\geq 0.15$, they have an opposite behavior to the 
cusp at $\epsilon=0.14$; $r_{\rm tr}$ increases at $T^{\ast}$ 
and drops just above $T^{\ast}$, and therefore have a different 
origin from the cusp related to the VS-VL phase transition. 

\begin{figure}[t]
\includegraphics[width=80mm]{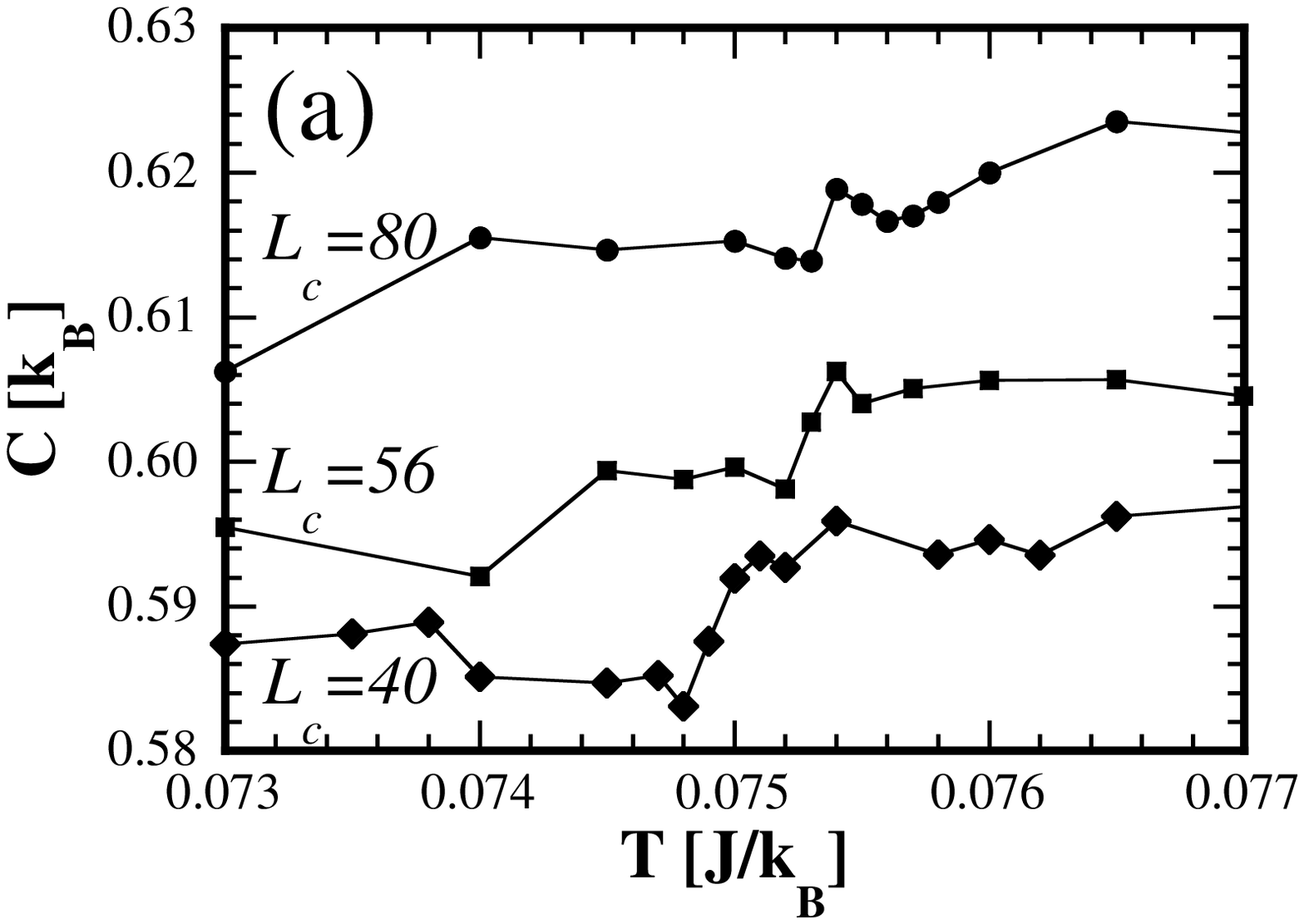}
\includegraphics[width=80mm]{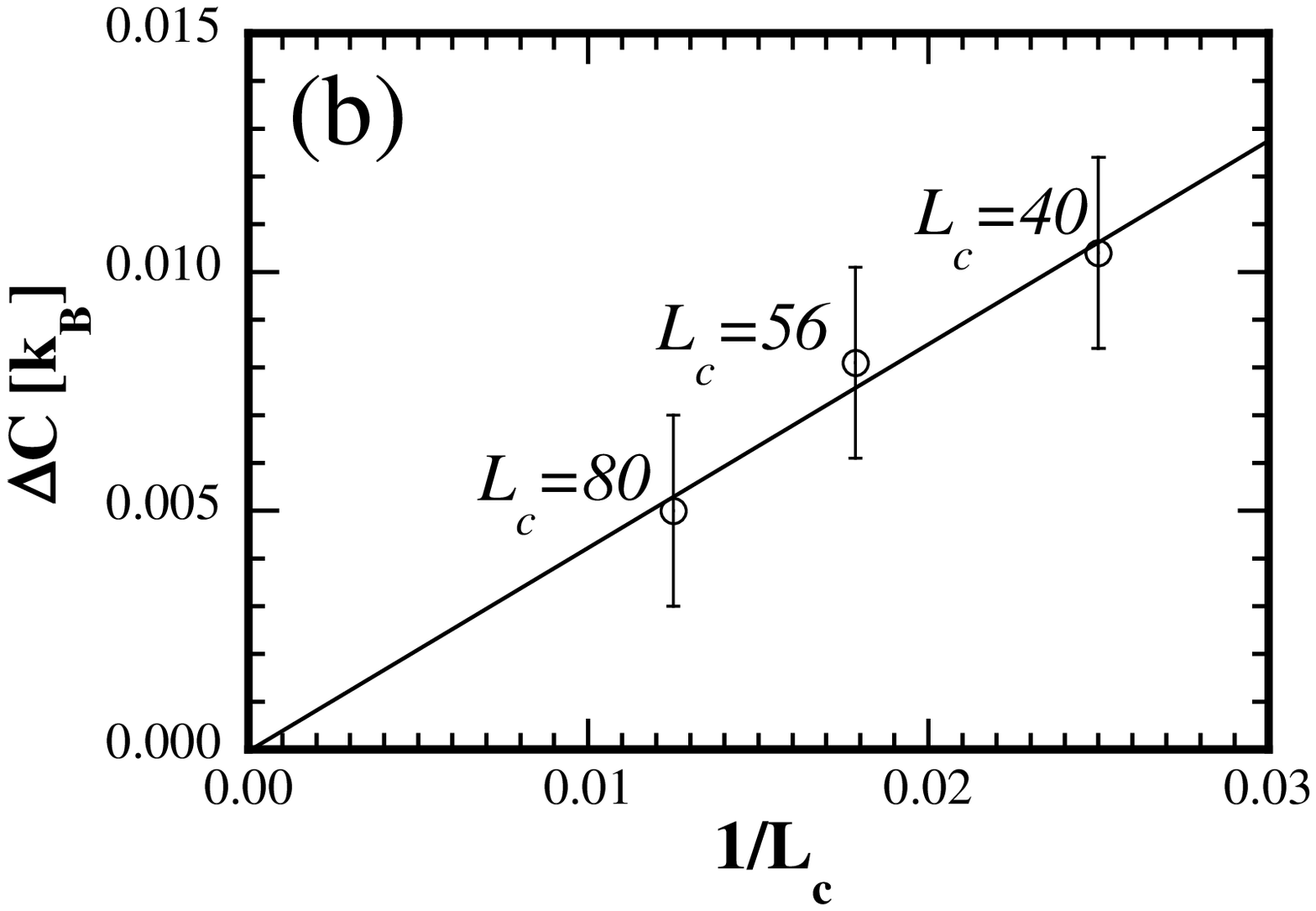}
\caption{\label{fig3}(a) Temperature dependence of the 
specific heat at $\epsilon=0.2$ for $L_{c}=40$, $56$ and 
$80$. Its origin is varied for each $\epsilon$. (b) $c$-axis 
size dependence of the height of the step in (a). The straight 
line is drawn by the least-squares fitting.
}
\end{figure}
Next, in order to explore whether this step-like anomaly 
of the specific heat is a thermodynamic phase transition 
or not, its size dependence is investigated. 
Since flux lines in ${\bm B} \parallel \hat{\bm c}$ are 
one-dimensional objects along the $c$ axis, physical 
quantities mainly depend on $L_{c}$. \cite{Nonomura2} 
Even in the vortex liquid region in pure systems, the 
displacement of flux lines along the $ab$ plane is 
smaller than the random-walk value $\sim L_{c}^{1/2}$ 
due to the blocking by other flux lines. In the present 
case, this displacement would be even smaller due to the 
pinning by defects. We thus fix $L_{a}=L_{b}=50$ and 
change $L_{c}$. Temperature dependence of the specific 
heat for $L_{c}=40$, $56$ and $80$ at $\epsilon=0.2$ is 
displayed in Fig.\ \ref{fig3}(a). As $L_{c}$ increases, the 
height of the step of the specific heat becomes smaller. Since 
it is clearly proportional to $1/L_{c}$ (Fig.\ \ref{fig3}(b)), 
this step is expected to vanish in the thermodynamic limit. 
The anomaly of the specific heat above the critical point 
of the VS-VL phase boundary is therefore a crossover 
rather than a thermodynamic phase transition. 

\begin{figure}[t]
\includegraphics[width=80mm]{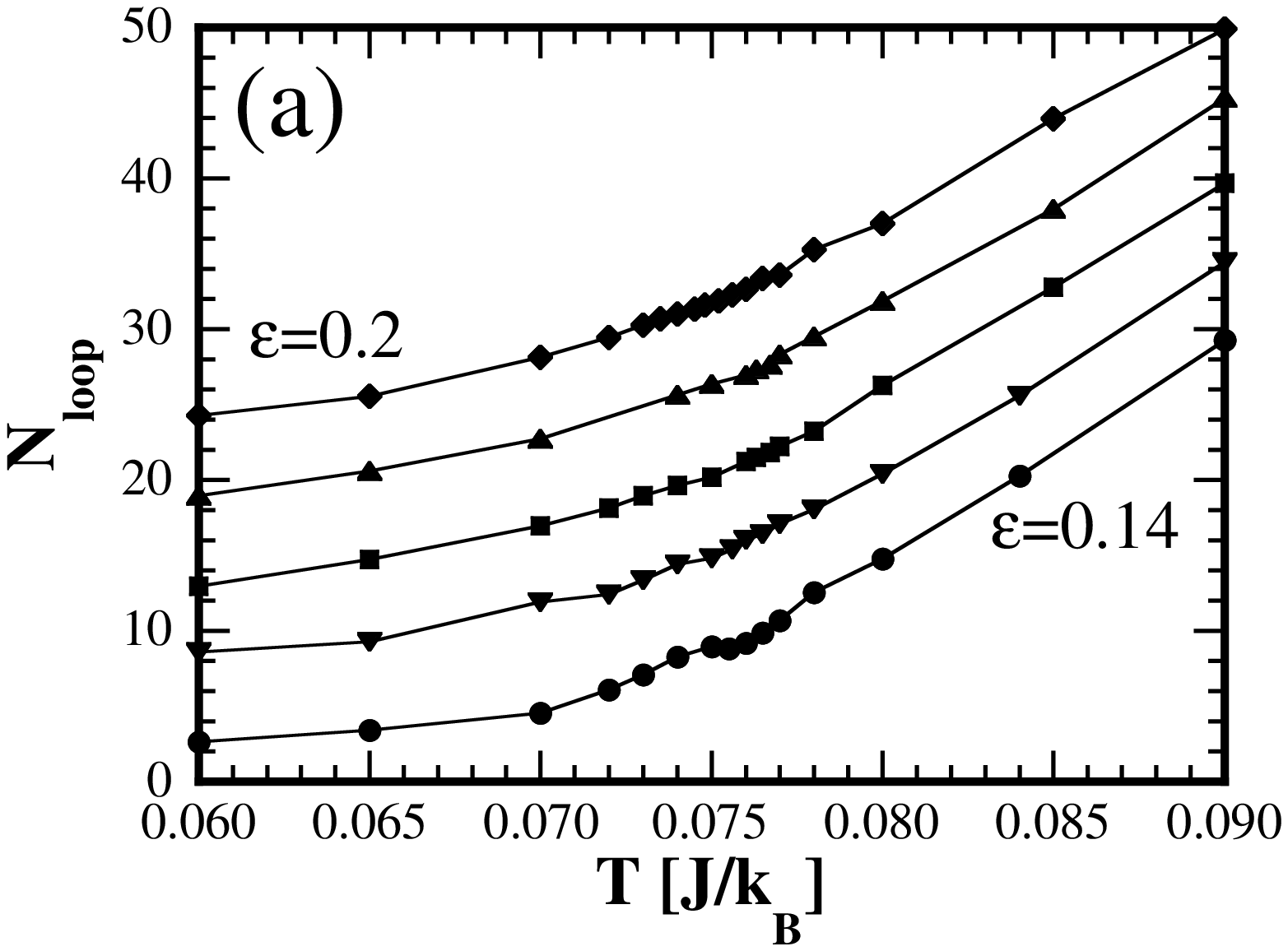}
\includegraphics[width=80mm]{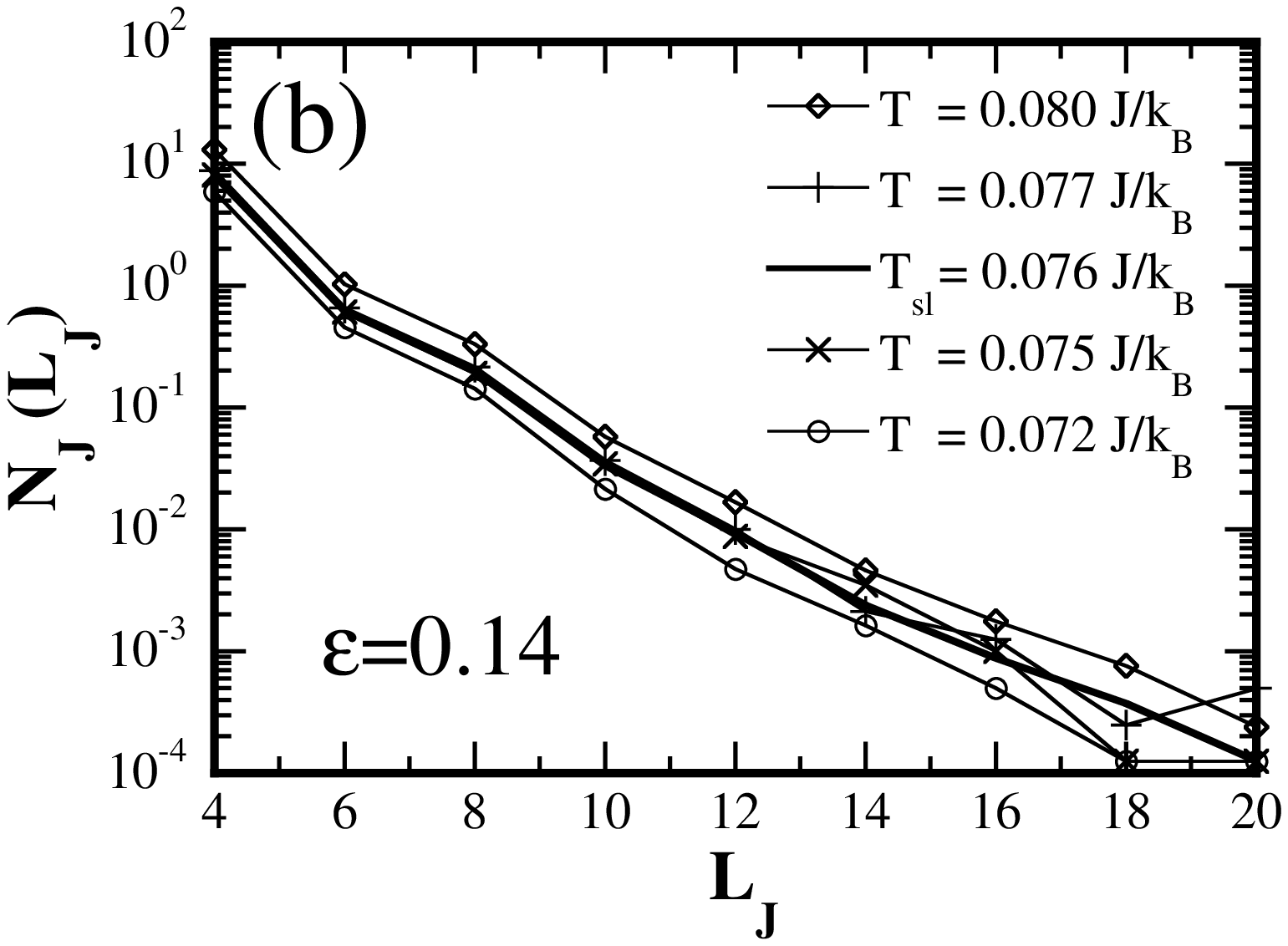}
\includegraphics[width=80mm]{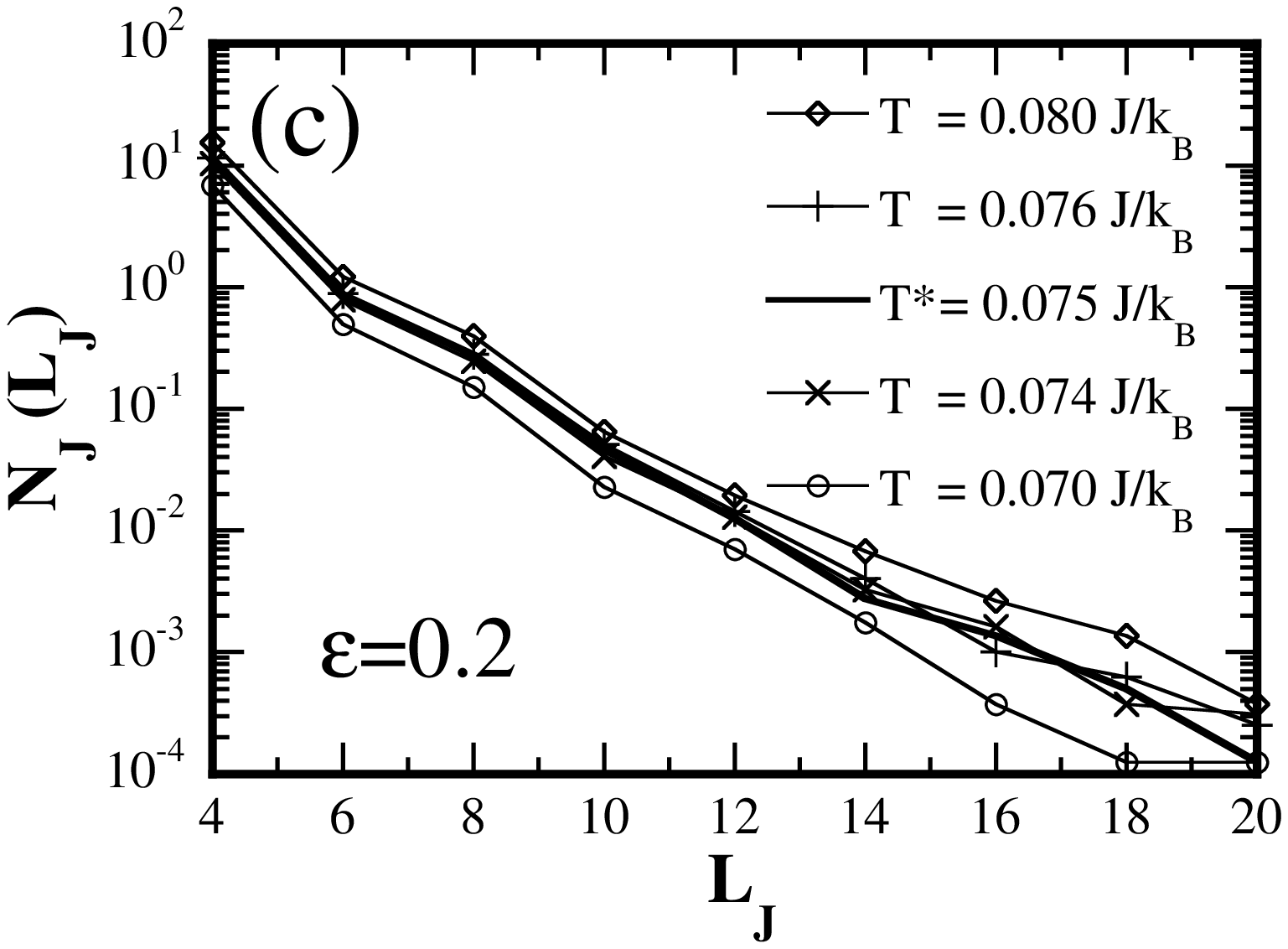}
\caption{\label{fig4}(a) Temperature dependence 
of the number of vortex loops for $\epsilon=0.14$, 
$0.15$, $0.16$, $0.18$ and $0.2$ (from bottom 
to top). Its origin is varied for each $\epsilon$. Size 
distributions of Josephson loops for (b) $\epsilon=0.14$ 
and (c) $\epsilon=0.2$ around the anomalies of 
the specific heat are drawn in a semi-log scale. 
}
\end{figure}
Finally, the number and size distribution of thermally-excited 
vortex loops are investigated in order to check whether this 
anomaly of the specific heat is related to the vortex-loop 
blowout \cite{Tesanovic,Nguyen} or not. 
Temperature dependence of the number of vortex loops for 
$0.14 \leq \epsilon \leq 0.2$ is displayed in Fig.\ \ref{fig4}(a). 
There exist no vortex loops including pancake vortices at these 
parameter regions. Size distributions of Josephson loops for 
$\epsilon=0.14$ and $0.2$ are shown in Figs.\ \ref{fig4}(b) 
and \ref{fig4}(c), respectively. \cite{Note}

Although the number of vortex loops monotonically increases as 
temperature increases, no symptom of the vortex-loop blowout is 
seen around $T_{\rm sl}$ or $T^{\ast}$ for each $\epsilon$. 
This is also the case for the size distribution of Josephson loops. 
In Figs.\ \ref{fig4}(b) and \ref{fig4}(c), the number of Josephson 
loops $N_{\rm J}(L_{\rm J})$ with a given perimeter $L_{\rm J}$ 
is plotted versus $L_{\rm J}$ in a semi-log scale, and the exponential 
size distribution of $N_{\rm J}(L_{\rm J})$ is clearly observed. 
There exists no qualitative difference in the behaviors below and above 
the critical point, and the size distribution changes continuously as 
temperature is varied around $T_{\rm sl}$ or $T^{\ast}$. 

The vortex-loop blowout is characterized by the abrupt increase of 
the number of vortex loops and the change of the size distribution 
from an exponential decay to a  power-law one at the transition 
temperature, and the proliferation of vortex loops as large as the 
system size above the transition temperature. \cite{Tesanovic,Nguyen} 
None of these behaviors are observed around $T^{\ast}$ 
in the present simulations, which casts serious doubt on the 
argument that the step of the specific heat is related to the 
vortex-loop blowout. \cite{Bouquet} In this argument, the 
vortex-loop blowout scenario was borrowed in order to explain 
the observation that the value of the specific heat above the 
step-like anomaly $C_{+}$ is larger than that below the anomaly 
$C_{-}$. However, this property is quite general in vortex states, 
and actually occurs even at the melting transition in pure 
systems. \cite{Hu} In vortex states, the density of vortices 
increases as temperature increases owing to excitation of 
vortex loops and wandering of flux lines, and consequently 
the relation $C_{+}>C_{-}$ holds. The vortex-loop blowout 
is merely a special case of this general property. 

It remains unsolved why a step of the specific heat is 
clearly observed in experiments even though the sample 
size is much larger than the system size in our simulations. 
Several possibilities are: (a) The height of the step of $C$ 
might saturate at larger $L_{c}$, and our maximum system 
size $L_{c}=80$ would be still small. (b) Our model (\ref{XYham}) 
might be still too simple to describe the step of $C$, e.g.\ 
the inhomogeneity of the strength of point defects is not 
taken into account. (c) The step of $C$ observed 
experimentally might be a non-equilibrium effect. 

In summary, vortex states of high-$T_{\rm c}$ superconductors 
with point defects in ${\bm B} \parallel \hat{\bm c}$ has been 
investigated by large-scale Monte Carlo simulations. The critical 
point of the first-order phase boundary between the vortex slush 
and vortex liquid phases is identified as the vanishing point of the 
$\delta$-function peak of the specific heat. The shape of the 
step-like anomaly of the specific heat above the critical point is 
similar to that of the experiment by Bouquet {\it et al.} As the 
system size along the $c$ axis $L_{c}$ increases, the height 
of this step seems proportional to $1/L_{c}$, which suggests 
that this anomaly is a crossover rather than a thermodynamic 
phase transition. Consistent behaviors are observed in the 
temperature dependence of the density of dislocations in the 
$ab$ plane and the ratio of point defects which trap flux lines. 
There is no discontinuity in the number and size distribution 
of vortex loops around the step-like anomaly of the specific 
heat, which clearly indicates that this anomaly is not related 
to the vortex-loop blowout. 

Numerical calculations were performed on Numerical Materials 
Simulator (NEC SX-5) at Computational Materials Science Center, 
National Institute for Materials Science, Japan. Y.~N.\ was 
supported by the Atomic Energy Research Fund from Ministry of 
Education, Culture, Sports, Science and Technology, Japan. 
This study is partially supported by the Priority Grant 
No.\ 14038240 from the same Ministry. 


\begin{thebibliography}{99}
\bibitem{Note}
The effect of intersections of vortex loops is negligible in the present parameter 
regions, because the averaged number of vortex loops in each $ab$ plane is 
less than unity, in contrast to the case of Ref.\ \cite{Olsson2}.
\end{thebibliography}
\end{document}